\documentclass[twocolumn]{aastex631}

\begin{document}

\title{AGN's Deadness Over Cosmic Time: $UVJ$ Diagrams of X-Ray-Selected AGN}

\author{Rachel Cionitti}
\affiliation{Department of Physics \& Astronomy, University of Kansas \\
1450 Jayhawk Blvd \\
Lawrence KS 66045, USA}

\author{Brandon Coleman}
\affiliation{Department of Physics \& Astronomy, University of Kansas \\
1450 Jayhawk Blvd \\
Lawrence KS 66045, USA}

\author{Allison Kirkpatrick}
\affiliation{Department of Physics \& Astronomy, University of Kansas \\
1450 Jayhawk Blvd \\
Lawrence KS 66045, USA}

\author{Greg Troiani}
\affiliation{Department of Physics \& Astronomy, University of Kansas \\
1450 Jayhawk Blvd \\
Lawrence KS 66045, USA}



\begin{abstract}
Active Galactic Nuclei (AGN) are intensely accreting supermassive black holes at the centers of massive galaxies. 
Though these objects occupy little spatial extent of the galaxy itself, they are thought to have far reaching affects, impacting the galaxy's star formation, and possibly 
it's lifespan until it becomes 'red and dead'. 
Typical galaxies 
demonstrate that, over cosmic time, they tend to 
separate into a bimodal distribution of 'red and dead' or blue and star forming. We examine whether active galaxies evolve over cosmic time in a similar way, and whether this can reveal anything about the complexities of the relationship between an AGN and 
the host galaxy.
We use the Stripe82X survey to identify 3940 X-ray AGN spanning $z=0-2.5$, and we measure the rest-frame $UVJ$ colors of each galaxy. We classify AGN as star-forming or quiescent based on their location in a $UVJ$ color diagram. We find that 
there is not a clear bimodal distribution between AGN in star forming and quiescent galaxies.
Furthermore, the most luminous X-ray sources tend to lie in the star forming region, which may indicate a correlation between central engine activity and increased rates of star formation.  

\end{abstract}



\section{Introduction} \label{sec:intro}

It is generally thought that active supermassive black holes at the center of galaxies impact the rest of the galaxy’s evolution and star formation rates through feedback, which is the mechanism by which the active center heats, cools or pressurizes the galaxy's gas content. However, the exact details of feedback are the subject of much debate 
\citep[for a review, see][]{Kormendy_2013}. Quasars are the brightest subclass of active galactic nuclei (AGN), so the effect of the feedback may be more pronounced, making them ideal objects for studying the relationship between nuclear activity and star formation rate. This AGN feedback on star formation can be positive or negative, depending on the ways in which the AGN is affecting the surrounding gas. Positive feedback occurs when activity around the central black hole causes outward shocks of radiation pressure, which compress gas and trigger star formation \citep[e.g.][]{Silk_2013}. Negative feedback is the opposite--activity in the nucleus causes winds that blow out {\color{black} the} excess gas and dust needed to generate new stars \citep{Shin_2019}; AGN radiation can also heat up gas in the interstellar medium, potentially suppressing the process of star formation \citep{McKinney_2021}. Past theoretical work has predicted that the star forming phase occurs early on in the quasar's life \citep{Hopkins_2006}. 
However, observational studies have shown that there is no clear correlation with star formation rate (SFR) and declining X-ray activity, where X-ray activity indicates the galactic nucleus is active \citep{Coleman_2022,Stanley_2015}. 


{\color{black} Galaxy SFR can be traced through the use of the $UVJ$ diagram \citep{Williams}. The 'colors' here are the $V-J$, or the difference of brightness between the $V$ and $J$ bands, and the $U-V$, which is identical but for the $U$ and $J$ bands. This color-color diagram combines rest-frame $V-J$ with $U-V$ to separate star-forming galaxies from quiescent galaxies, which have quenched their star formation. These diagrams have been shown to be effective to $z\sim2$ \citep{Williams}. As the stellar populations initially age in a galaxy, the $U$-band emission will weaken first. This causes a galaxy to move upward in $UVJ$ colorspace. Increasing the age of the population will then begin to affect the $V-J$ color, causing galaxies to move to the right in colorspace. Dust can also affect the $V-J$ color, causing galaxies to appear redder and further to the right.

In this work, we explore how AGN strength correlates with star formation through the use of the $UVJ$ diagram and X-ray luminosities of our sample. In Section 2, we describe our survey data from Stripe82X, and discuss how the rest-frame $UVJ$ photometry is calculated. In Section 3, we show how X-ray luminosity correlates with location in $UVJ$ space from $z=0-2.5$. We present our Conclusions in Section 4. 
In this analysis, we assume a standard cosmology of H$_0$ = 70 Mpc/km/s, $\Lambda$ = 0.7, and $\Omega_m$ = 0.3. 

\section{Data \& Analysis}
\subsection{Sample Selection}
The data in this work comes from the Accretion History of AGN survey\footnote{https://project.ifa.hawaii.edu/aha/}. 
This survey contains X-ray observations of {\color{black} $>$6000} galaxies from the Stripe 82X survey \citep{LaMassa_2016}. The X-ray data covers 31.3 deg$^2$ of the \textit{Sloan Digital Sky Survey {\color{black} (SDSS)}} Stripe 82 Legacy Field, with observations coming from \textit{\color{black} the X-ray Multi-Mirror Mission} \textit{(XMM-Newton)} and {\color{black} the \textit{Chandra X-ray Observatory}}. 
The {\color{black} X-ray survey} data 
{\color{black} has been} crossmatched with many other multi-wavelength observations for a final sample of 6181 sources \citep{Ananna_2017}. Of the 6181 sources {\color{black} detected at X-ray wavelengths}, 6044 have corresponding multi-wavelength associations. {\color{black} For this work, we make use of $ugriz$ photometry from SDSS and $JHK$ photometry from the United Kingdom Infrared Telescope (UKIRT) Infrared
Deep Sky Survey (UKIDSS).}

{\color{black} For Stripe82X, spectroscopic redshifts are available from SDSS for 53\% of our sources. For the remainder of the galaxies, photometric redshifts were measured in \citet{Ananna_2017} by fitting a suite of templates to the optical photometry.} 
These templates 
are empirical and include 
starburst galaxies, Type 1 and Type 2 AGN, quasars, Seyferts, spirals and ellipticals. {\color{black} There are 206 galaxies which are not well-fit with any of the templates and which do not have a measured spectroscopic redshift. We remove these from the sample.}

We {\color{black} further} select all galaxies with observations in at least two bands {\color{black} from the $ugriz$ and $JHK$ photometry for a sample of 5514. We limit our analysis to $z<2.5$, as this is the redshift range for which the $UVJ$ diagram has been calibrated \citep{Williams}. There are 5180 of these sources with $0<z<2.5$, where 60\% are spectroscopic redshifts. These galaxies span a hard X-ray luminosity range of $L_{2-10\,{\rm keV}}=2.7*10^{34}-7.7*10^{45}$erg/s. 3029 have $L_{2-10\,{\rm keV}}\geq10^{42}$erg/s, which is the typical threshold for declaring a source an AGN. For the remainder of this work, $L_X$ refers to the hard X-ray luminosity, $L_{2-10\,{\rm keV}}$.}






\subsection{Measuring Rest Frame UVJ Photometry}

{\color{black} We} fit the 
visible and infrared (\textit{ugriz}, \textit{JHK}) 
photometry for each source {\color{black} with the suite of empirical templates from \citep{Ananna_2017} after shifting all photometry to the rest-frame and converting magnitudes to fluxes.} 

{\color{black} We use the following equation to fit the templates:}

\begin{equation}
    f = N* L_{\nu,template}*\lambda^\alpha
\end{equation}
{\color{black} $f$ represents the SED template after it has been fitted to best match the observed fluxes. $N$ and $\alpha$ are free parameters. $N$ must be positive, but $\alpha$ is allowed to be positive, zero, or negative. This form can allow for a more precise fitting than simply a scaling factor alone. We are exclusively concerned with achieving the best fit possible to the data, to allow for the most accurate calculation of the rest-frame $UVJ$ photometry. As such, we are not concerned with a physical interpretation of $\alpha$, nor do we draw any inference from the best-fit template (which can be star-forming, elliptical, Seyfert, etc.).}
{\color{black} We select the best-fitting model by calculating an $R^2$ parameter for each fit. The $R^2$ statistic measures the ratio of the variance in the residuals (in this case, measured photometry vs.\,model-predicted photometry) to the variance in the data (photometry). The model with $R^2$ closest to 1 is then used to calculate the $UVJ$ photometry.}

{\color{black} We convolve $UVJ$ transmission curves with each best fit model to calculate synthetic, rest-frame photometry for each galaxy. We use the Bessel transmission curves which have central wavelengths of 3543 \AA, 5510 \AA, and 12000 \AA $ $ for the $U, V$, and $J$ filters, respectively.}
We visually inspected the best fit template, galaxy photometry, and new synthetic photometry to ensure the results were consistent with the real galaxy observations. 

\section{Results \& Discussion}

\begin{figure*}
    \centering
    \includegraphics[width=1.0\linewidth]{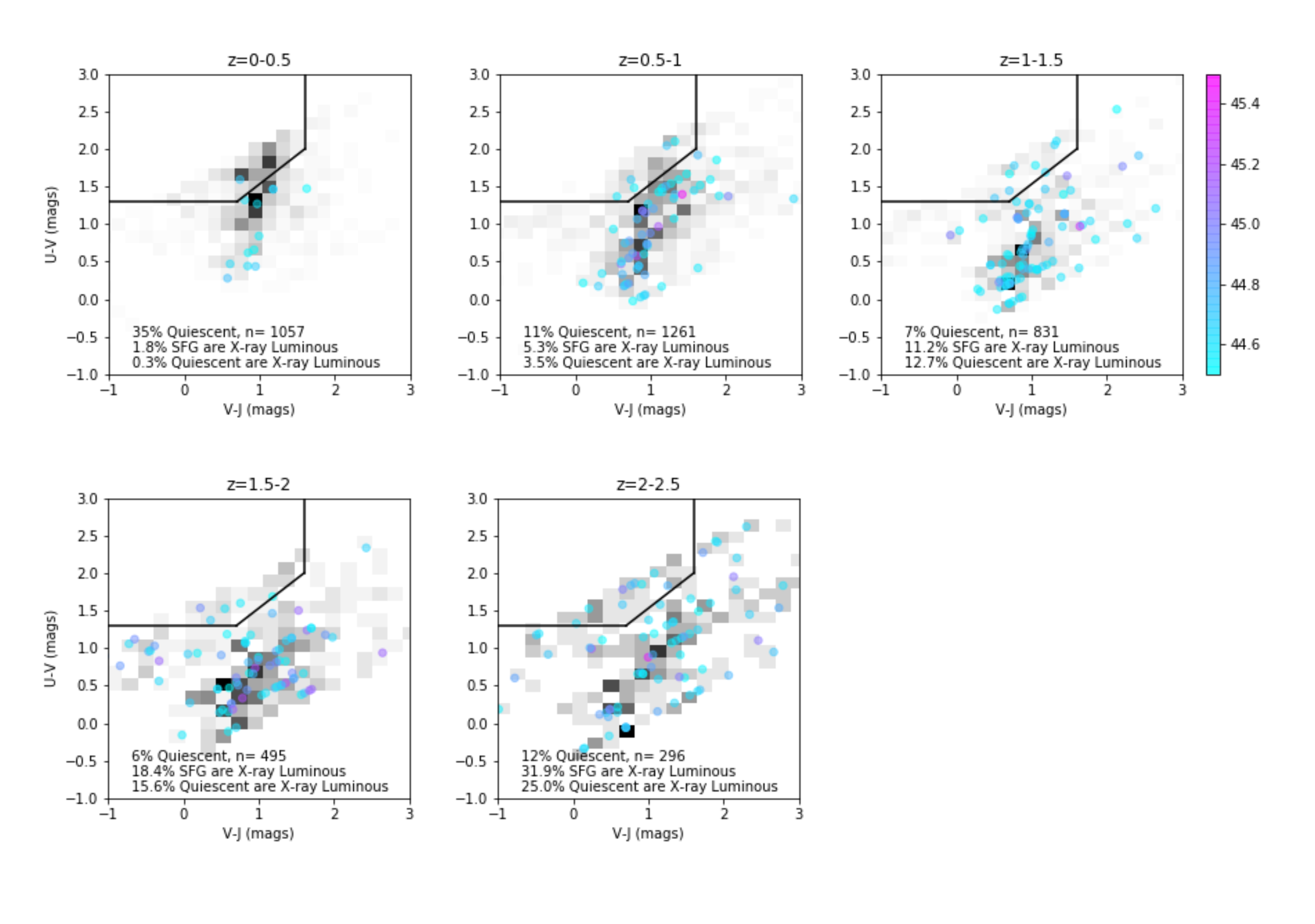}
    \caption{The $UVJ$ diagrams for all objects in the sample at each redshift bin. {\color{black} The quiescent region falls in the upper left and is indicated by the solid black line.} The black and white is a 2D histogram representing the density of sources at each region. The colored points represent the exact locations of the most luminous X-ray sources {\color{black} ($\log L_X > 44.5$).} 
    The color bar demonstrates intensity in log(erg/s) in the hard X-ray band. {\color{black} In each panel, we list the number of sources and the percentage of quiescent galaxies. We also list the percentage of galaxies in the star forming region that are X-ray luminous and the percentage in the quiescent region that are X-ray luminous.}} 
    \label{fig:xray}
\end{figure*}

Figure \ref{fig:xray} shows the $UVJ$ plots {\color{black} of our sources using the synthetic rest-frame photometry.} 
The sources are split by redshift intervals of $\Delta z=0.5$. The quiescent {\color{black} region} is set by 
\cite{Williams}.
The black and white portions on the plot represent a 2D histogram of the density of sources in each position in color space, where the darker the bin is, the more populated that region is. {\color{black} The lower redshift bins contain $>$1000 sources each, while the highest redshift bin has only $\sim$300 galaxies. Each panel in Figure \ref{fig:xray} indicates the quiescent fraction of X-ray sources at each epoch. In general}, the quiescent fraction increases 
{\color{black} with decreasing redshift}, similar to typical galaxies (\cite{Williams}, \cite{Madau_2014}). {\color{black} At $z<0.5$, 35\% of X-ray sources are quiescent, compared with 6\% at $z=1.5-2.0$. This number increases to 12\% at $z>2$, but the location of the quiescent region grows more uncertain at higher redshift. This is due to two reasons. First, star-forming galaxies at $z>2$ are dustier as gas fractions increase \citep[e.g.,][]{whitaker2017}, which could contaminate the right edge of the $UVJ$ quiescent region. Second, the calculation of rest-frame photometry becomes more uncertain as it relies heavily on extrapolation (the $K$-band corresponds to $\lambda_{\rm rest}=7000$\, \AA $ $ at $z=2.0$, which is a lower wavelength than the rest-frame $J$ photometry that we are estimating.}

{\color{black} Our sources span a range of X-ray luminosities and therefore could include non-AGN, which could bias the results. To test this, we consider the location of only the brightest sources with $L_X>10^{44.5}$\,erg/s.}
The highly luminous X-ray sources are indicated by the shaded dots. {\color{black} There is no clear trend of X-ray luminous sources preferentially residing in the quiescent region, as might be expected if quasar activity comes after star formation.}
In general, the highly X-ray luminous sources tend to be slightly biased toward residing in the star forming region, indicative of positive feedback. 

{\color{black} Notably, neither the full X-ray population nor the X-ray luminous sources demonstrate as clear of a bimodal distribution as seen in non-AGN. This could be due to a significant portion of AGN residing in the ``green valley'', the transition region in colorspace between star forming and quiescent galaxies \citep[e.g.,][]{mahoro2017}. The position of AGN in $UVJ$ space may indicate that AGN do not preferentially reside in any type of galaxy; AGN can be found in star forming, quiescent, or transitioning galaxies. This non-preference could be due to the timescales of AGN variability making a correlation between AGN luminosity and host galaxy properties difficult to idenfity. It could also be due to the complicated relationship between AGN and host, with feedback both quenching and triggering star formation.} 

 This apparent star forming bias for the highly X-ray luminous sources (and also for the full population) indicates that there may be a correlation between central engine activity, which is signalled by extreme X-ray luminosity, and star formation in the host galaxy. This is interesting because it is suggesting a sort of positive feedback, where central engine activity may cause phenomena such as shocks that can compress the surrounding gas and increase star formation. {\color{black} This result is in line with what is seen in \citet{kocevski2017} where the bulk of X-ray-selected AGN at $z\sim2$ in the EGS and UDS fields lie in galaxies that are still star forming, but are more compact than normal star forming galaxies. However, this skew of the highly luminous sources toward the star forming region of the diagram is not obvious or especially strong.}

{\color{black} Finally, we note that it is possible the AGN can contaminate the optical portion of a galaxy's spectral energy distribution,} so the $UVJ$ is tracing the accretion disk, not the galaxy. {\color{black} However, X-ray emission will mimic elliptical galaxies most strongly, so we anticipate that the quiescent fractions will be even lower if AGN emission is a major source of contamination.} 

\section{Conclusion}

 This work set out to assess 
 {\color{black} whether} X-ray selected AGN {\color{black} were preferentially star forming} using an empirical $UVJ$ color-color diagram, which 
 indicates {\color{black} whether a source is star forming or quiescent}
 {\color{black}. We calculated rest-frame $UVJ$ colors for 3940 X-ray AGN from $z=0-2.5$ and classified galaxies as quiescent based on their location in $UVJ$ colorspace}. The results indicate that active galaxies trend toward quiescence {\color{black} with decreasing redshift, similar to normal star forming galaxies.} 
 However, {\color{black} we also find} 
 that there does not seem to be a clear correlation between central engine activity {\color{black} as probed by X-ray luminosity} and the location of the sources on the $UVJ$ diagram. There is a near ubiquitous trend where the highly X-ray luminous sources tend to be more star forming than quiescent, but this is not by a significant amount. However, the possibility of a correlation can not be ruled out either, and can be honed with further studies. {\color{black} This work shows that X-ray AGN seem to reside in star forming galaxies, quiescent galaxies, and those transitioning to a quiescent phase.} 

\begin{acknowledgments}
\end{acknowledgments}

%




\bibliography{sample631}{}

\begin{thebibliography}{}
\expandafter\ifx\csname natexlab\endcsname\relax\def\natexlab#1{#1}\fi
\providecommand{\url}[1]{\href{#1}{#1}}
\providecommand{\dodoi}[1]{doi:~\href{http://doi.org/#1}{\nolinkurl{#1}}}
\providecommand{\doeprint}[1]{\href{http://ascl.net/#1}{\nolinkurl{http://ascl.net/#1}}}
\providecommand{\doarXiv}[1]{\href{https://arxiv.org/abs/#1}{\nolinkurl{https://arxiv.org/abs/#1}}}

\bibitem[{Ananna {et~al.}(2017)Ananna, Salvato, LaMassa, Urry, Cappelluti,
  Cardamone, Civano, Farrah, Gilfanov, Glikman, Hamilton, Kirkpatrick,
  Lanzuisi, Marchesi, Merloni, Nandra, Natarajan, Richards, \&
  Timlin}]{Ananna_2017}
Ananna, T.~T., Salvato, M., LaMassa, S., {et~al.} 2017, The Astrophysical
  Journal, 850, 66, \dodoi{10.3847/1538-4357/aa937d}

\bibitem[{Coleman {et~al.}(2022)Coleman, Kirkpatrick, Cooke, Glikman, Massa,
  Marchesi, Peca, Treister, Auge, Urry, Sanders, Turner, \&
  Ananna}]{Coleman_2022}
Coleman, B., Kirkpatrick, A., Cooke, K.~C., {et~al.} 2022, Monthly Notices of
  the Royal Astronomical Society, 515, 82.
\newblock \url{https://doi.org/10.1093/mnras/stac1679}

\bibitem[{Hopkins \& Hernquist(2006)}]{Hopkins_2006}
Hopkins, P.~F., \& Hernquist, L. 2006, The Astrophysical Journal Supplement
  Series, 166, 1, \dodoi{10.1086/505753}

\bibitem[{{Kocevski} {et~al.}(2017){Kocevski}, {Barro}, {Faber}, {Dekel},
  {Somerville}, {Young}, {Williams}, {McIntosh}, {Georgakakis}, {Hasinger},
  {Nandra}, {Civano}, {Alexander}, {Almaini}, {Conselice}, {Donley},
  {Ferguson}, {Giavalisco}, {Grogin}, {Hathi}, {Hawkins}, {Koekemoer}, {Koo},
  {McGrath}, {Mobasher}, {P{\'e}rez Gonz{\'a}lez}, {Pforr}, {Primack},
  {Santini}, {Stefanon}, {Trump}, {van der Wel}, {Wuyts}, \&
  {Yan}}]{kocevski2017}
{Kocevski}, D.~D., {Barro}, G., {Faber}, S.~M., {et~al.} 2017, \apj, 846, 112,
  \dodoi{10.3847/1538-4357/aa8566}

\bibitem[{Kormendy \& Ho(2013)}]{Kormendy_2013}
Kormendy, J., \& Ho, L.~C. 2013, Annual Review of Astronomy and Astrophysics,
  51, 511, \dodoi{10.1146/annurev-astro-082708-101811}

\bibitem[{LaMassa {et~al.}(2016)LaMassa, Urry, Cappelluti, Böhringer,
  Comastri, Glikman, Richards, Ananna, Brusa, Cardamone, Chon, Civano, Farrah,
  Gilfanov, Green, Komossa, Lira, Makler, Marchesi, Pecoraro, Ranalli, Salvato,
  Schawinski, Stern, Treister, \& Viero}]{LaMassa_2016}
LaMassa, S.~M., Urry, C.~M., Cappelluti, N., {et~al.} 2016, The Astrophysical
  Journal, 817, 172, \dodoi{10.3847/0004-637X/817/2/172}

\bibitem[{Madau \& Dickinson(2014)}]{Madau_2014}
Madau, P., \& Dickinson, M. 2014, Annual Review of Astronomy and Astrophysics,
  52, 415, \dodoi{10.1146/annurev-astro-081811-125615}

\bibitem[{{Mahoro} {et~al.}(2017){Mahoro}, {Povi{\'c}}, \&
  {Nkundabakura}}]{mahoro2017}
{Mahoro}, A., {Povi{\'c}}, M., \& {Nkundabakura}, P. 2017, \mnras, 471, 3226,
  \dodoi{10.1093/mnras/stx1762}

\bibitem[{{McKinney} {et~al.}(2021){McKinney}, {Hayward}, {Rosenthal},
  {Mart{\'\i}nez-Galarza}, {Pope}, {Sajina}, \& {Smith}}]{McKinney_2021}
{McKinney}, J., {Hayward}, C.~C., {Rosenthal}, L.~J., {et~al.} 2021, \apj, 921,
  55, \dodoi{10.3847/1538-4357/ac185f}

\bibitem[{Shin {et~al.}(2019)Shin, Woo, Chung, Baek, Cho, Kang, \&
  Bae}]{Shin_2019}
Shin, J., Woo, J.-H., Chung, A., {et~al.} 2019, The Astrophysical Journal, 881,
  147, \dodoi{10.3847/1538-4357/ab2e72}

\bibitem[{{Silk}(2013)}]{Silk_2013}
{Silk}, J. 2013, \apj, 772, 112, \dodoi{10.1088/0004-637X/772/2/112}

\bibitem[{Stanley {et~al.}(2015)Stanley, Harrison, Alexander, Swinbank, Aird,
  Moro, Hickox, \& Mullaney}]{Stanley_2015}
Stanley, F., Harrison, C.~M., Alexander, D.~M., {et~al.} 2015, Monthly Notices
  of the Royal Astronomical Society, 453, 591, \dodoi{10.1093/mnras/stv1678}

\bibitem[{{Whitaker} {et~al.}(2017){Whitaker}, {Pope}, {Cybulski}, {Casey},
  {Popping}, \& {Yun}}]{whitaker2017}
{Whitaker}, K.~E., {Pope}, A., {Cybulski}, R., {et~al.} 2017, \apj, 850, 208,
  \dodoi{10.3847/1538-4357/aa94ce}

\bibitem[{Williams {et~al.}(2009)Williams, Quadri, Franx, van Dokkum, \&
  Labb{\'{e} }}]{Williams}
Williams, R.~J., Quadri, R.~F., Franx, M., van Dokkum, P., \& Labb{\'{e} }, I.
  2009, The Astrophysical Journal, 691, 1879,
  \dodoi{10.1088/0004-637x/691/2/1879}

\end{thebibliography}
\bibliographystyle{aasjournal}



\end{document}